\documentclass[aps,pre,onecolumn,groupedaddress,longbibliography,nofootinbib,showkeys,showpacs]{revtex4-1}
\usepackage{graphicx}
\usepackage{mathptmx}      
\usepackage{bm}
\usepackage{enumerate}
\usepackage{subfigure}
\usepackage{color}

\DeclareRobustCommand\openone{\leavevmode\hbox{\small1\normalsize\kern-.33em1}}
\newcommand\bsigma{{\bm\sigma}}
\usepackage{enumerate}

\begin{document}
\title{Discrete-event simulation of uncertainty in single-neutron experiments\footnote{
Published in Front. Physics 2:14. doi: 10.3389/fphy.2014.00014}
}

\author{Hans De Raedt}
\affiliation{Department of Applied Physics, Zernike Institute for Advanced Materials,\\
University of Groningen, Nijenborgh 4, NL-9747AG Groningen, The Netherlands}
\author{Kristel Michielsen}
\thanks{Corresponding author}
\email{k.michielsen@fz-juelich.de}
\affiliation{Institute for Advanced Simulation, J\"ulich Supercomputing Centre,\\
Forschungszentrum J\"ulich, D-52425 J\"ulich, Germany}
\affiliation{RWTH Aachen University, D-52056 Aachen, Germany}

\begin{abstract}
A discrete-event simulation approach which provides a cause-and-effect description of many
experiments with photons and neutrons exhibiting interference and entanglement is applied to a recent
single-neutron experiment that tests (generalizations of) Heisenberg's uncertainty relation.
The event-based simulation algorithm reproduces the results of the
quantum theoretical description of the experiment
but does not require the knowledge of the solution of a wave equation nor does it
rely on concepts of quantum theory.
In particular, the data satisfies uncertainty relations derived in the context of quantum theory.
\end{abstract}


\maketitle

\section{Introduction}
Quantum theory has proven extraordinarily powerful for describing a vast number of laboratory experiments.
The mathematical framework of quantum theory allows for a straightforward (at least in principle)
calculation of numbers which can be compared with experimental
data as long as these numbers refer to statistical
averages of measured quantities, such as for example atomic spectra,
the specific heat and magnetic susceptibility of solids.
However, as soon as an experiment is able to record the individual
clicks of a detector which contribute to the statistical average a fundamental problem appears.
Although quantum theory provides a recipe to compute
the frequencies for observing events, it does not account
for the observation of the individual events themselves~\citep{BALL70,HOME97,BALL03,NIEU13}.
Prime examples are the single-electron two-slit experiment~\citep{TONO98},
single-neutron interferometry experiments~\citep{RAUC00} and
optics experiments in which the click
of a detector is assumed to correspond to the arrival of a single photon~\citep{GARR08}.

From the viewpoint of quantum theory, the central issue is how it can be that
experiments yield definite answers.
On the other hand, it is our brain which decides, based on what it perceives through our senses and cognitive capabilities,
what a definite answer is and what it is not.
According to Bohr~\citep{BOHR99}
``Physics is to be regarded not so much as the study of something a priori given,
but rather as the development of methods of ordering and surveying human experience.
In this respect our task must be to account for such experience in a manner independent
of individual subjective judgment and therefore objective in the sense that it can be
unambiguously communicated in ordinary human language''.
This quote may be read as a suggestion to construct a description in terms of events,
some of which are directly related to human experience, and the cause-and-effect relations among them.
Such an event-based description obviously yields definite answers and if
it reproduces the statistical results of experiments,
it also provides a description on a level to which quantum theory has no access.

For many interference and entanglement phenomena observed in optics and neutron experiments,
such an event-based description has already been constructed, see ~\citep{MICH11a,RAED12a,RAED12b} for recent reviews.
The event-based simulation models reproduce the statistical distributions of quantum theory
without solving a wave equation but by modeling physical phenomena as a chronological sequence of events.
Hereby events can be actions of an experimenter, particle emissions by a source, signal generations by a detector,
interactions of a particle with a material and so on~\citep{MICH11a,RAED12a,RAED12b}.

The basic premise of our event-based simulation approach is that
current scientific knowledge derives from the discrete events which
are observed in laboratory experiments and from relations between those events.
Hence, the event-based simulation approach is concerned with
how we can model these experimental observations but not with what ``really'' happens in Nature.
This underlying premise strongly differs from the assumption that the observed events are signatures
of an underlying objective reality which is mathematical in nature
but is in line with Bohr's viewpoint expressed in the above quote.

The general idea of the event-based simulation method is that simple rules define
discrete-event processes which may lead to the behavior that is observed in experiments.
The basic strategy in designing these rules is to carefully examine
the experimental procedure and to devise rules such that they produce the same kind of data as those
recorded in experiment, while avoiding the trap of simulating thought experiments that are difficult to realize in
the laboratory. Evidently, mainly because of the lack of knowledge, the rules are not unique.
Hence, it makes sense to use the simplest rules one could think of until a new experiment indicates that the rules should be modified.
The method may be considered as entirely ``classical'' since it only uses concepts
which are directly related to our perception of the macroscopic world
but the rules themselves are not necessarily those of classical Newtonian dynamics.

The event-based approach has successfully been used for discrete-event simulations of
quantum optics experiments such as
the single beam splitter and
Mach-Zehnder interferometer experiments, 
Wheeler's delayed choice experiments, 
a quantum eraser experiment, 
two-beam single-photon interference experiments and the single-photon interference experiment with
a Fresnel biprism, 
Hanbury Brown-Twiss experiments, 
Einstein-Podolsky-Rosen-Bohm (EPRB) experiments, 
and of conventional optics problems such as the propagation of electromagnetic plane waves through
homogeneous thin films and stratified media, see \citep{MICH11a,RAED12a} and references therein. 
For applications to single-neutron interferometry experiments see~\citep{RAED12a,RAED12b}.
The same methodology has also been employed to perform discrete-event simulations of
quantum cryptography protocols~\citep{ZHAO08a} and universal quantum computation~\citep{MICH05}.

In this paper, we extend this list by demonstrating that the same approach provides
an event-by-event description of recent neutron experiments~\citep{ERHA12,SULY13} devised to test (generalizations of)
Heisenberg's uncertainty principle.
It is shown that the event-by-event simulation generates data which complies with the quantum theoretical description
of this experiment.
Therefore, these data also satisfy the inequalities which, in quantum theory, express
(generalizations of) Heisenberg's uncertainty principle.
However, as the event-by-event simulation does not resort to concepts of quantum theory
these findings indicate that there is little intrinsically ``quantum mechanical''
to these inequalities, in concert with the idea
that quantum theory can be cast into a ``classical'' statistical theory~\citep{FRIE89,REGI98,HALL00,FRIE04,KHRE09,KHRE11,KHRE11a,KAPS10,SKAL11,KAPS11,KLEI12,KLEI12a,FLEG12}.

\section{Experiment and quantum theoretical description}\label{sec2}

\begin{figure}[t]
\begin{center}
\includegraphics[width=\hsize]{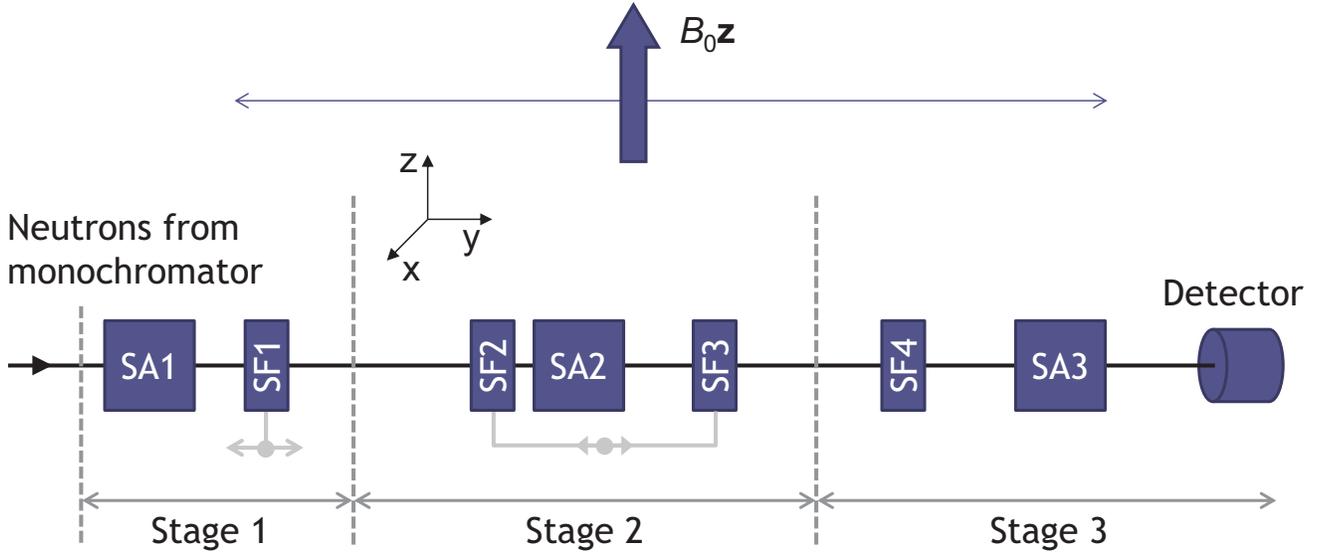}
\end{center}
\caption{Functional block diagram of the neutron experiment~\citep{ERHA12,SULY13} to test
uncertainty relations, see also Fig.~2 in~\citep{ERHA12,SULY13}.
Monochromatic neutrons enter from the left.
SA1, SA2, SA3: spin analyzers (see text);
SF1, SF2, SF3, SF4: spin flippers (see text).
The positions of SF1 and the pair (SF2,SF3) are variable.
The preparation stage 1 manipulates the magnetic moments of the neutrons such that,
depending on the orientation of SA1 and the position of SF1, they leave stage 1
with their moments directed along the $z$-axis or along a fixed direction
on the $x-y$ plane.
In the quantum theoretical description (see text), stage 2 and stage 3
perform the successive measurement of the eigenvalues of the
spin operators $\sigma_\phi=\sigma_x \cos\phi+\sigma_y\sin\phi$ and $\sigma_y$, respectively.
These seven devices, in combination with the static magnetic field $B_0\mathbf{z}$ and the
variable positions of SF1 and (SF2,SF3), are sufficient
to determine the expectation values that appear in Heisenberg-like uncertainty relations (see text).
}
\label{stages}
\end{figure}

A block diagram of the neutron experiment designed to test uncertainty relations~\citep{ERHA12,SULY13}
is shown in Fig.~\ref{stages}.
We now describe this experiment in operational terms and as we go along, we also give
the quantum theoretical description in terms of spin 1/2 particles such as neutrons.

Conceptually, the neutron experiment~\citep{ERHA12,SULY13} exploits two different physical phenomena:
the motion of a magnetic moment in a static magnetic field and a spin-analyzer that performs
a Stern-Gerlach-like selection of the neutrons based on the direction of their magnetic moments.

A magnetic moment $\mathbf{S}$ in an external, static magnetic field $B\mathbf{e}$ experiences a rotation about
the direction of the unit vector $\mathbf{e}$.
The unitary transformation that corresponds to such a rotation is given by
\begin{eqnarray}
U_\mathbf{b}(\varphi)&=&
e^{i\gamma t B \mathbf{S}\cdot\mathbf{e}}
=
e^{i\varphi\bsigma\cdot\mathbf{e}}
,
\label{qt0}
\end{eqnarray}
where $\gamma$ is the gyromagnetic ration of the particle,
$t$ is the time that the particle interacts with the magnetic field,
the variable $\varphi=\gamma t B$ is introduced as a shorthand
for the angle of rotation,
and $\bsigma=(\sigma_x,\sigma_y,\sigma_z)$ denote the Pauli-spin matrices.

The spin analyzer acts as a projector.
It is a straightforward (see pages 172 and 250 in \citep{BALL03})
to show that within quantum theory, an ideal spin analyzer directed along
the unit vector ${\mathbf n}$ is represented by the projection operator
\begin{eqnarray}
M(S,\bm{n})&=& \frac{\openone+S \bsigma\cdot\mathbf{n}}{2}
,
\label{qt1}
\end{eqnarray}
where $\openone$ is the unit matrix and
$S=\pm1$ selects one of the two possible alignments of the spin polarizer along ${\mathbf n}$
(see \citep{ERHA12,SULY13}).

Using Eqs.~(\ref{qt0}) and (\ref{qt1}), it is straightforward to construct
the quantum theoretical description of
each of the three stages in the experimental setup.

\begin{enumerate}[{\bf Stage} \bf 1.\ ]
\item%
The purpose of the first spin analyzer (SA1) is to prepare neutrons with their magnetic moments
in the direction of the static magnetic field $B_0 \boldsymbol{z}$.
Then, the particle travels for some time in a region where the field $B_0 \boldsymbol{z}$ is present
but as its magnetic moment is aligned along $z$,
the magnetic moment does not rotate.

As will become clear later, to test Ozawa's inequality~\citep{OZAW03,ERHA12,SULY13}, it is necessary to be able
to prepare initial states in which the magnetic moment lies in the $x-y$ plane.
In the experiment, this is accomplished by putting in place, the spin flipper SF1.
The spin flipper (SF1), in essence a static magnetic field aligned along the $x$-direction,
rotates the magnetization about the $x$-axis by an amount proportional to the
magnetic field. For simplicity, it is assumed that this rotation changes
the direction of the magnetic moment from $z$ to $y$~\citep{ERHA12,SULY13}.
The position of SF1, relative to the direction of flight of the neutrons, is variable.
By moving SF1 one can change the time that the particles perform rotations about the $z$-axis,
hence one can control the direction of the magnetic moment in the $x-y$ plane as it leaves stage 1.

Quantum theoretically, the action of the components of stage 1 is described by the product of unitary matrices
\begin{eqnarray}
U_1&=&U_{\mathbf{z}}(\theta_1)U_{\mathbf{x}}(\theta_0)
,
\label{qt2}
\end{eqnarray}
where  $\theta_0=\pi/2$ or $\theta_0=0$ if SF1 is in place or not and $\theta_1$ is the variable
(through the variable position of SF1) rotation angle, the value of which will be fixed later.
Obviously, in the case that SF1 is not present, because the incoming neutrons have their moments
aligned along the $z$-direction, these moments do not perform rotations at all.
\item%
This stage consists of a pair of spin flippers (SF2,SF3) and a spin analyzer (SA2).
The position of (SF2,SF3), relative to the direction of flight of the neutrons, is variable
whereas the position of SA2 is fixed.
The action of the components of stage 2 is described by the product of matrices
\begin{eqnarray}
T_2&=&U_{\mathbf{z}}(\theta_4)U_{\mathbf{x}}(\pi/2)
M(S_1,\bm{z})
U_{\mathbf{z}}(\theta_3)U_{\mathbf{x}}(\pi/2)U_{\mathbf{z}}(\theta_2)
,
\label{qt3}
\end{eqnarray}
where, as a consequence of the variable position of (SF2,SF3), the rotation angles
$\theta_2$, $\theta_3$, and $\theta_4$ change with the position of (SF2,SF3).
The value of variable $S_1=\pm1$ labels one of the two possible alignments of the spin polarizer along $\mathbf{z}$.
Note that because of the projection $M(S_1,\bm{z})$, the matrix $T_2$ is not unitary.
\item%
The final stage consists of a spin flipper SF4 and a spin analyzer SA3.
The time evolution of the magnetic moment as it traverses this stage is given by
\begin{eqnarray}
T_3&=&M(S_2,\bm{z})U_{\mathbf{z}}(\theta_5)U_{\mathbf{x}}(\pi/2)
,
\label{qt4}
\end{eqnarray}
where $\theta_5$ is a fixed rotation angle.
The value of variable $S_2=\pm1$ labels one of the two possible alignments of the spin polarizer along $\mathbf{z}$.
The matrix $T_3$ is not unitary.
\end{enumerate}
According to the postulates of quantum theory, the probability to detect a neutron leaving stage 3
is given by~\citep{BALL03}
\begin{eqnarray}
P(S_1,S_2|\rho)&=&\mathbf{Tr} \rho(T_3T_2U_1)^\dagger (T_3T_2U_1)
,
\label{qt5}
\end{eqnarray}
where it is assumed that the detector simply counts every impinging neutron (which in view
of the very high detection efficiency is a very good approximation, see ~\citep{RAUC00}).
In Eq.~(\ref{qt5}), the initial state of the S=1/2 quantum system is represented by the density matrix
\begin{eqnarray}
\rho&=&\frac{\openone+\bsigma\cdot\mathbf{a}}{2}
.
\label{qt6}
\end{eqnarray}
The real-valued vector $\mathbf{a}$ ($\Vert\mathbf{a}\Vert\le1$) completely determines the initial state of the quantum system.
Using Eq.~(\ref{qt2}) -- (\ref{qt6}) we find
\begin{eqnarray}
P(S_1,S_2|\mathbf{a})&=&
\frac{1 + (S_1 - S_2 \cos\theta_4) (a_x \sin(\theta_1+\theta_2) - a_y \cos(\theta_1+\theta_2))- S_1S_2 \cos\theta_4 }{4}
,
\label{qt7}
\end{eqnarray}
independent of $\theta_3$ and $\theta_5$.
Recall that by construction of the experimental setup $\theta_2+\theta_4$ is fixed.
We can make contact to the expressions used in the analysis of the neutron experiment~\citep{ERHA12,SULY13},
by substituting $\theta_1=0$, $\theta_2=\phi+\pi/2$ and $\theta_4=-\phi-\pi/2$
where $\phi$ is called the detuning angle~\citep{ERHA12,SULY13}.
We obtain
\begin{eqnarray}
P(S_1,S_2|\mathbf{a})&=&
\frac{1 + (S_1 + S_2 \sin\phi) (a_x \cos\phi + a_y \sin\phi)+  S_1S_2 \sin\phi}{4}
.
\label{qt8}
\end{eqnarray}
As explained in detail in subsection~\ref{filter}, the experimental setup can be interpreted
as performing successive measurements of the operators
$\sigma_\phi=\sigma_x \cos\phi + \sigma_y \sin\phi$
and $\sigma_y$, their eigenvalues being $S_1$ and $S_2$, respectively.
Note that these two operators do not commute unless $\cos\phi=0$
and that the observed eigenvalues $S_1$ and $S_2$ of these two operators are correlated,
as is evident from the contribution $S_1S_2\sin\phi$ in Eq.~(\ref{qt8}).

From Eq.~(\ref{qt8}), the expectation values of the various spin operators
follow immediately. Specifically, we have

\begin{eqnarray}
\langle \sigma_\phi \rangle_\mathbf{a}&=&\sum_{S_1,S_2=\pm1} S_1 P(S_1,S_2|\mathbf{a})
=a_x \cos\phi + a_y \sin\phi,
\nonumber \\
\langle \sigma_y \rangle_\mathbf{a}&=&\sum_{S_1,S_2=\pm1} S_2 P(S_1,S_2|\mathbf{a})
=\sin\phi(a_x \cos\phi + a_y \sin\phi)=\sin\phi\langle \sigma_\phi \rangle_\mathbf{a}
,
\label{qt9}
\end{eqnarray}
and as $\sigma_\phi^2=\sigma_y^2=1$, the variances
$\langle \sigma_\phi^2 \rangle_\mathbf{a}-\langle \sigma_\phi \rangle_\mathbf{a}^2$ and
$\langle \sigma_y^2 \rangle_\mathbf{a}-\langle \sigma_y \rangle_\mathbf{a}^2$
are completely determined by Eq.~(\ref{qt9}).

\subsection{Filtering-type measurements of one spin-1/2 particle}\label{filter}

\begin{figure}[t]
\begin{center}
\includegraphics[width=12cm]{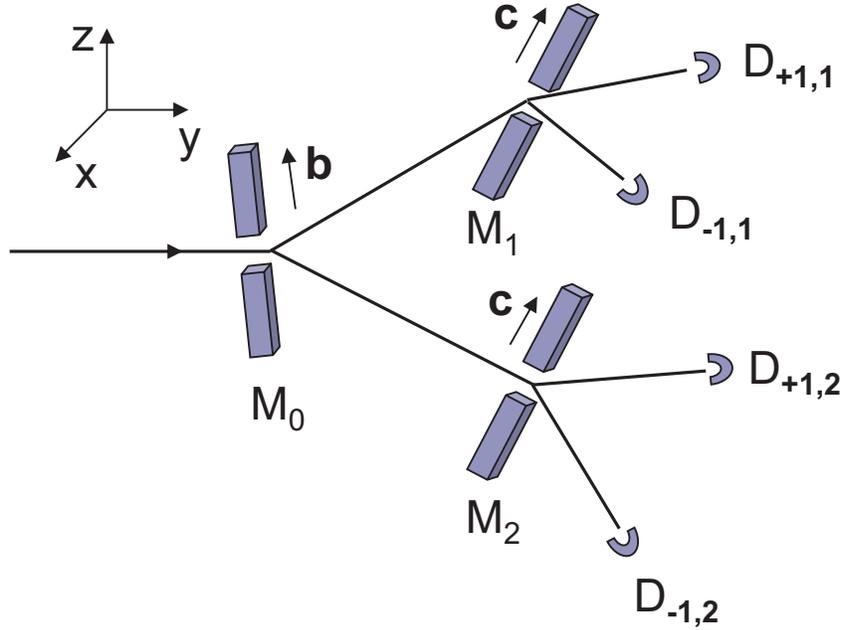}
\end{center}
\caption{Diagram of the filtering-type experiment of which
the neutron experiment~\citep{ERHA12,SULY13} is a special case.
Spin-1/2 particles pass through a Stern-Gerlach magnet $M_0$
that projects the spin onto either the
${\mathbf b}$ direction or the $-{\mathbf b}$ direction.
In case of the former (latter) projection, the particle
is directed to the Stern-Gerlach magnet $M_1$ ($M_2$).
$M_1$ and $M_2$ are assumed to be identical and
project the spin onto either the
${\mathbf c}$ direction or the $-{\mathbf c}$ direction.
A ``click'' of one of the four detectors
$D_{+1,1}$, $D_{-1,1}$, $D_{+1,2}$, and $D_{-1,2}$
signals the arrival of a particle.
}
\label{fig0}
\end{figure}

The neutron experiment~\citep{ERHA12,SULY13} can be viewed as a particular realization of a filtering-type
experiment~\citep{BALL03,RAED11a}.
The layout of such an experiment is shown in Fig.~\ref{fig0}.
According to this diagram, the experiment consists of performing successive Stern-Gerlach-type measurements
on one spin-1/2 particle at a time.
Conceptually, assuming a stationary particle source, the neutron experiment~\citep{ERHA12,SULY13} and the
filtering-type experiment shown in Fig.~\ref{fig0} are identical, see also Fig.~1 in~\citep{ERHA12,SULY13}.
In practice, the difference between the filtering-type experiment and the neutron experiment~\citep{ERHA12,SULY13}
is that in the latter four experiments (labeled by the variables $S_1=\pm1$ and $S_2=\pm1$),
are required for each of the two opposite orientations of the two spin analyzers SA2 and SA3
whereas the setup depicted in Fig.~\ref{fig0} directly yields the results of the four separate runs.

The pictorial description of the filtering experiment goes as follows.
A particle enters the Stern-Gerlach magnet $M_0$, with its
magnetic field along direction ${\mathbf b}$. $M_0$ ``sends" the particle
either to Stern-Gerlach magnet $M_1$ or $M_2$. The magnets
$M_1$ and $M_2$, identical and both with their magnetic field along
direction ${\mathbf c}$, redirect the particle once more and
finally, the particle is registered by one of the four
detectors $D_{+1,1}$, $D_{-1,1}$, $D_{+1,2}$, and $D_{-1,2}$.
This scenario is repeated until the statistical fluctuations
of the counts of the four detectors are considered to be sufficiently small.

We label the particles by a subscript $\alpha$. After the $\alpha$th
particle leaves $M_1$ or $M_2$, it will hit one (but only one) of the four
detectors. We assume ideal experiments, that is at any time one and
only one out of four detectors fires. We write $x_\alpha^{(i,j)}=1$
with $i=\pm1$ and $j=1,2$
if the $\alpha$th particle was detected by detector $D_{i,j}$ and
$x_\alpha^{(i,j)}=0$ otherwise.

We define two new dichotomic variables by
\begin{eqnarray}
S_{1,\alpha}&=&
\left( x_\alpha^{(+1,1)}+x_\alpha^{(-1,1)}\right)
-
\left( x_\alpha^{(+1,2)}+x_\alpha^{(-1,2)}\right)
,
\nonumber \\
S_{2,\alpha}&=&
\left( x_\alpha^{(+1,1)}+x_\alpha^{(+1,2)}\right)
-
\left( x_\alpha^{(-1,1)}+x_\alpha^{(-1,2)}\right)
.
\label{fil0}
\end{eqnarray}
Note that for each incoming particle, only one of the detectors clicks
hence only one of the $x_\alpha^{(i,j)}$'s is nonzero or,
equivalently
$x_\alpha^{(+1,1)}+x_\alpha^{(-1,1)}+x_\alpha^{(+1,2)}+x_\alpha^{(-1,2)}=1$.

In the quantum theoretical description of the filtering experiment,
if $S_{1,\alpha}=\pm1$, the spin has been projected on the
$\pm{\mathbf b}$ direction. Likewise, if $S_{2,\alpha}=\pm1$, the
spin has been projected on the $\pm{\mathbf c}$ direction.
In other words, $S_{1,\alpha}$  and $S_{2,\alpha}$
are the eigenvalues of the spin operator projected on
the directions ${\mathbf b}$ and ${\mathbf c}$, respectively.
Then, according to quantum theory, the probability to observe a pair of eigenvalues
$(S_{1},S_{2})$ is given by~\citep{BALL03,RAED11a}
\begin{eqnarray}
P(S_{1},S_{2}|\mathbf{a})&=&
\mathbf{Tr} \rho \left(M(S_2,{\mathbf c})M(S_1,{\mathbf b})\right)^\dagger  \left(M(S_2,{\mathbf c})M(S_1,{\mathbf b})\right)
\nonumber \\&=&
\mathbf{Tr} \rho M(S_1,\mathbf{b})M(S_2,{\mathbf c})M(S_1,{\mathbf b})
\nonumber \\&=&
\frac{1+(S_1+ \mathbf{b}\cdot\mathbf{c}\;S_2)\mathbf{a}\cdot\mathbf{b}
+\mathbf{b}\cdot\mathbf{c}\;S_1 S_2}{4}
,
\label{fil6}
\end{eqnarray}
where the state $\rho$ is given by Eq.~(\ref{qt6}) and the $M$'s denote projection operators.
It is easy to see that Eq.~(\ref{qt8}) is a particular case of Eq.~(\ref{fil6}).

Note that $[M(S_1,{\mathbf b}),M(S_2,{\mathbf c})]\not=0$ unless
$\mathbf{b}=\pm\mathbf{c}$, $[\rho,M(S_1,{\mathbf b})]\not=0$ unless
$\mathbf{a}=\pm\mathbf{b}$, and that $[\rho,M(S_2,{\mathbf c})]\not=0$ unless $\mathbf{a}=\pm\mathbf{c}$.
Thus, for virtually all cases of interest, none of the operators in Eq.~(\ref{fil6})
commute, yet quantum theory yields the probability $P(S_1,S_2|\mathbf{a})$ for all cases.
Clearly, the statement that one can determine the eigenvalues of two non-commuting
operators in one experiment contradicts the conventional teaching that
non-commuting operators cannot be diagonalized simultaneously
and {\bf therefore} cannot be measured simultaneously.
The reason for this apparent contradiction is
the hidden assumption that diagonalization and the act of measurement
in a laboratory (i.e. a click of the detector) are equivalent in some sense.
The filtering-type experiment is a clear example which shows that they are not:
according to quantum theory the eigenvalues
$S_1$ and $S_2$ of the operators $\bsigma\cdot\mathbf{b}$ and $\bsigma\cdot\mathbf{c}$, respectively
can always be measured simultaneously even though these operators cannot always be diagonalized simultaneously.

\section{Event-by-event simulation}

A minimal, discrete-event simulation model of single-neutron experiments requires a specification of the
information carried by the neutrons, of the algorithm that simulates the source
and the devices used in the experimental setup (see Fig.~\ref{stages}), and of the procedure to analyze the data.

\begin{itemize}
\item
{\bf Messenger:}
A neutron is regarded as a messenger carrying a message.
In principle, there is a lot of freedom to specify the content of the message, the only
criterion being that in the end, the simulation should reproduce the results of real laboratory experiments.
We adopt Occam's razor as a guiding principle to determine which kind of data the messenger should carry with it, that
is we use the minimal amount of data.

The pictorial description that will be used in the following should not be taken literally:
it is only meant to help visualize, in terms of concepts familiar from macroscopic physics,
the minimal amount of data the messenger should carry.

Picturing the neutron as a tiny classical magnet
we can use the spherical coordinates $\theta$ and $\varphi$
to specify the direction of its magnetic moment
\begin{equation}
{\mathbf m}=(\cos\varphi\sin \theta, \sin\varphi\sin \theta, \cos \theta)^T
,
\label{eve1}
\end{equation}
relative to the fixed frame of reference defined by the static magnetic field $B_0\mathbf{z}$.
The messenger should also be aware of the time it takes to move from one point in space to another.
The time of flight and the direction of the magnetic moment
are conveniently encoded in a message of the type~\citep{RAED12a,RAED12b}
\begin{equation}
{\mathbf u}=(e^{i\psi^{(1)}}\cos (\theta/2), e^{i\psi^{(2)}}\sin (\theta/2))^T
,
\label{neutron}
\end{equation}
where
$\psi^{(i)} =\nu t +\delta_i$, for $i=1,2$
and $\varphi=\delta_1-\delta_2=\psi^{(1)}-\psi^{(2)}$.
Within the present model, the state of the neutron, that is the message, is completely described
by the angles $\psi^{(1)}$, $\psi^{(2)}$ and $\theta$ and by rules (to be specified),
by which these angles change as the neutron travels through the network of devices.
This model suffices to reproduce the results of single-neutron interference and entanglement
experiments and of their idealized quantum theoretical descriptions~\citep{RAED12a,RAED12b}.

In specifying the message Eq.~(\ref{neutron}),
we exploited the isomorphism between the algebra of Pauli matrices and rotations in three-dimensional space,
not because the former connects to quantum mechanics
but only because we find this representation most convenient for our simulation work~\citep{MICH11a,RAED12a,RAED12b}.
The direction of the magnetic moment follows from Eq.~(\ref{neutron}) through
\begin{equation}
{\mathbf m}={\mathbf u}^T \bsigma {\mathbf u}
.
\label{moment}
\end{equation}

A messenger with message ${\mathbf u}$ at time $t$ and position ${\mathbf r}$
that travels with velocity $v$, along the direction ${\mathbf q}$ during a time interval
$t^{\prime} - t$, changes its message according to $\psi^{(i)}\leftarrow \psi^{(i)}+\nu(t^{\prime}-t)$ for $i=1,2$,
where $\nu$ is an angular frequency which is characteristic for a neutron
that moves with a fixed velocity $v$.
In a monochromatic beam of neutrons, all neutrons have the same value of $\nu$~\citep{RAUC00}.

In the presence of a magnetic field ${\mathbf B}=(B_x,B_y,B_z)$, the magnetic moment
rotates about the direction of ${\mathbf B}$ according to the
classical equation of motion
\begin{equation}
\frac{d \mathbf{m}}{dt}=\mathbf{m}\times{\mathbf B}
.
\label{eve2}
\end{equation}
Hence, as the messenger passes a region in which a magnetic field is present,
the message ${\mathbf u}$ changes into the message
\begin{equation}
{\mathbf u}\leftarrow e^{ig\mu_N T\mathbf {\sigma}\cdot {\mathbf B} /2} {\mathbf u}
,
\label{eve3}
\end{equation}
where
$g$ denotes the neutron $g$-factor, $\mu_N$ the nuclear magneton,
$T$ the time during which the messenger experiences the magnetic field.

In the event-based simulation of the experiment shown in Fig.~\ref{stages},
the time-of-flight $T$ determines the angle of rotation of the magnetic moment through Eq.~(\ref{eve3}) and
can, so to speak, be eliminated by expressing all operations in terms of rotation angles.
However, this simplification is no longer possible in the event-based simulation of single-neutron
interference and entanglement experiments~\citep{RAED12a,RAED12b}.
\item
{\bf Source:}
When the source creates a messenger, its message needs to be initialized.
This means that three angles $\psi^{(1)}$, $\psi^{(2)}$ and $\theta$ need to be specified.
In practice, instead of implementing stage 1 it is more efficient to prepare the messengers
such that the corresponding magnetic moments are along a specified, fixed direction.
For instance, to mimic fully coherent, spin-polarized neutrons that enter stage 2 with their spin along the $x$-axis,
the source would create messengers with $\theta=\pi/2$, and without loss of generality, $\psi^{(1)}=\psi^{(2)}=0$.
\item
{\bf Spin-flipper:}
The spin-flipper rotates the magnetic moment by an angle $\pi/2$ about the $x$ axis.
\item
{\color{black}
{\bf Spin analyzer:}
This device shares with the magnet used in Stern-Gerlach experiments the property that it
diverts incoming particles in directions which depend on the orientation of their magnetic moments
relative to the magnetic field inside the device.
For appropriate choices of the experimental parameters, the Stern-Gerlach magnet
splits the incoming beam of particles into spatially separated beams.
If there are two well-separated beams, the action of the device its to align the
magnetic moments of the incoming particles
either parallel or anti-parallel to the direction of the field.

Ignoring all the details of interaction of the magnetic moments with the Stern-Gerlach magnet,
the operation of separating the incoming beam into spatially separated beams
is captured by the very simple probabilistic model defined by
\begin{equation}
x=2\Theta\left(\frac{1+m_zS}{2}-r\right)-1
,
\label{eve4}
\end{equation}
where $x=-1,1$ labels the two distinct spatial directions, $0<r<1$ is a uniform pseudo-random number,
$\Theta(x)$ is the unit step function and,
as before, $S=\pm1$ labels the orientation of the spin analyzer.
For each incoming messenger, a new pseudo-random number is being generated.
Recall that $|m_z|\le1$ (see Eq.~(\ref{eve1})) hence the first term of Eq.~(\ref{eve4}) is
a number between zero and two. If we would set $m_z = \langle\bsigma\cdot\mathbf{n}\rangle$,
the model defined by Eq.~(\ref{eve4}) would generate minus- and plus ones according
to the projection operator Eq.~(\ref{qt1}).
Applied to the neutron experiments~\citep{ERHA12,SULY13}, the function of the spin analyzer
is to pass particles with say, spin-up, only.
In the simulation model this translates to letting the messenger pass if
$x=1$ and destroy the messenger if $x=-1$.

It is of interest to explore the possibility whether different models for the spin analyzer
can yield averages over many events which cannot be distinguished
from the results obtained by employing the probabilistic model Eq.~(\ref{eve4}).
As the extreme opposite to the probabilistic model, we consider a deterministic
learning machine (DLM) defined by the update rule
\begin{eqnarray}
x&=&2\Theta\left(m_zS- \gamma u\right)-1
\nonumber \\
u&\leftarrow& \gamma u + (1-\gamma)u
,
\label{eve4a}
\end{eqnarray}
where $x=-1,1$ labels the two distinct spatial directions
and $-1\le u\le 1$ encodes the internal state of the DLM
(the equivalent of the seed of the pseudo-random number generator).
The parameter $0\le\gamma<1$ controls the pace at which the DLM learns the value $m_zS$.
The properties of the time series of $x$'s generated by the rules Eq.~(\ref{eve4a})
have been scrutinized in great detail elsewhere, see ~\citep{MICH11a} and references therein.
Suffice it to say here that for many events,
the average of the $x$'s converges to $m_zS$ and that
the $x$'s are highly correlated in time.

Obviously, the DLM-based model is extremely simple and fully deterministic.
It may easily be rejected as a viable candidate model by comparing
the correlations in experimentally observed time series with
those generated by Eq.~(\ref{eve4a}).
However, if the experiment only provides data about average quantities,
there is no way to rule out the DLM model.
Unfortunately, the neutron experiments~\citep{ERHA12,SULY13}
do not provide the data necessary to reject the DLM model,
simply because the spin analyzer passes particles with say, spin-up, only.
Hence there is no way to compute time correlations.
Although we certainly do not want to suggest that the spin analyzers
used in the experiments behave in the extreme deterministic manner
as described by Eq.~(\ref{eve4a}),
it is of interest to test whether such a simple deterministic model
can reproduce the averages computed from quantum theory
and also obeys the same uncertainty relations as the genuine quantum
theoretical model.

}
\item
{\bf Detector:}
As a messenger enters the detector, the detection count is increased by one and the messenger is destroyed.
The detector counts all incoming messengers.
Hence, we assume that the detector has a detection efficiency of 100\%.
This is a good model of real neutron detectors which can have a detection efficiency of $99\%$
or more~\citep{KROU00}.
\item
{\bf Simulation procedure and data analysis:}
First, we establish the correspondence between the initial message
$\mathbf{u}_{\mathrm{initial}}$ and the description
in terms of the density matrix Eq.~(\ref{qt6}).
To this end, we remove all devices from stage 1 and 2 and simply count the number of messages that pass SA3
with $S_2=1$, for instance.
It directly follows from Eq.~(\ref{eve4}) that the relative frequency of counts is given by $m_z$,
the projection of the message onto the $z$-axis.
In other words, we would infer from the data that
in a quantum theoretical description the $z$-component of the density matrix $a_z$ is equal to $m_z$.
By performing rotations of the original message
it follows by the same argument that $\mathbf{a}=\mathbf{m}_{\mathrm{initial}}$.

For each pair of settings $(S_1,S_2)$ of the spin analyzers (SA2,SA3)
and each position of the pair of spin flippers (SF2,SF3) represented by a rotation of $\phi$ about the $z$-axis
(see Section~\ref{sec2}), the source sends $N$ messengers through
the network of devices shown in Fig.~\ref{stages}.
The source only creates a new messenger if (i) the previous messenger has been processed by the detector or (ii)
the messenger was destroyed by one of the spin analyzers.
In other words, direct communication between messengers is excluded.
As a device in the network receives a messenger, it processes the message according to the rules specified earlier
and sends the messengers with the new message to the next device in the network.
If the device is a spin analyzer, it may happen that the messenger is destroyed.
The detector counts all messengers that pass SA3 and destroys these messengers.

For a sequence of $N$ messengers all carrying the same initial message $\mathbf{a}=\mathbf{m}_{\mathrm{initial}}$,
this procedure yields a count $N(S_1,S_2|\mathbf{a})$ (recall that $\phi$ is fixed during each sequence of $N$ events).
Repeating the procedure for the four pairs of settings yields the relative frequencies
\begin{eqnarray}
F(S_1,S_2|\mathbf{a})
&=&
\frac{
N(S_1,S_2|\mathbf{a})
}{
\sum_{S_1,S_2=\pm1} N(S_1,S_2|\mathbf{a})
}
.
\label{eve5}
\end{eqnarray}
Note that the numerator in Eq.~(\ref{eve5}) is not necessarily equal to $N$ because messengers may be
destroyed when they enter a spin analyzer.
From Eq.~(\ref{eve5}) we compute
\begin{eqnarray}
\langle S_1\rangle&=&\sum_{S_1,S_2=\pm1} S_1 F(S_1,S_2|\mathbf{a})
\label{eve5a}
,
\\
\langle S_2\rangle&=&\sum_{S_1,S_2=\pm1} S_2 F(S_1,S_2|\mathbf{a})
\label{eve5b}
,
\\
\langle S_1 S_2\rangle&=&\sum_{S_1,S_2=\pm1} S_1 S_2 F(S_1,S_2|\mathbf{a})
.
\label{eve5c}
\end{eqnarray}
\item
{\bf Validation.}
The event-based model reproduces the results of the quantum theoretical description if,
within the usual statistical fluctuations, we find that
$F(S_1,S_2|\mathbf{a})\approx P(S_1,S_2|\mathbf{a})$
with $P(S_1,S_2|\mathbf{a})$ given by Eq.~(\ref{qt8}).
This correspondence is most easily established by noting that for fixed $\phi$ and $\mathbf{a}$,
the three expectations Eqs.~(\ref{eve5a})--(\ref{eve5c}) completely determine Eq.~(\ref{eve5})
and that, likewise,  the quantum theoretical distribution Eq.~(\ref{qt8}) is
completely determined by the expectations Eqs.~(\ref{eve5a})--(\ref{eve5c}) with
$F(S_1,S_2|\mathbf{a})$ replaced by $P(S_1,S_2|\mathbf{a})$.
In other words, for the event-based model to reproduce the results of
the quantum theoretical description of the neutron experiment~\citep{ERHA12,SULY13},
it is necessary and sufficient that the simulation results for Eqs.~(\ref{eve5a})--(\ref{eve5c})
are in agreement with the quantum theoretical results  (see Eq.~(\ref{qt9}))
$\langle S_1\rangle=a_x \cos\phi + a_y \sin\phi$,
$\langle S_2\rangle=\sin\phi \langle S_1\rangle$,
and $\langle S_1 S_2\rangle=\sin\phi$.

{\color{black}
In Figs.~\ref{res1} and \ref{res1a}, we present representative results of event-based simulations
of the neutron experiment~\citep{ERHA12,SULY13}, showing that the simulation indeed reproduces
the predictions of the quantum theoretical description of the neutron experiment~\citep{ERHA12,SULY13}.
Comparing Figs.~\ref{res1} and \ref{res1a}, we can only conclude that it does not matter
whether the model for the spin analyzers uses pseudo-random numbers (Eq.~(\ref{eve4})) or
is DLM-based (Eq.~(\ref{eve4a})).
}

\end{itemize}

\begin{figure}[t]
\begin{center}
\includegraphics[width=8cm]{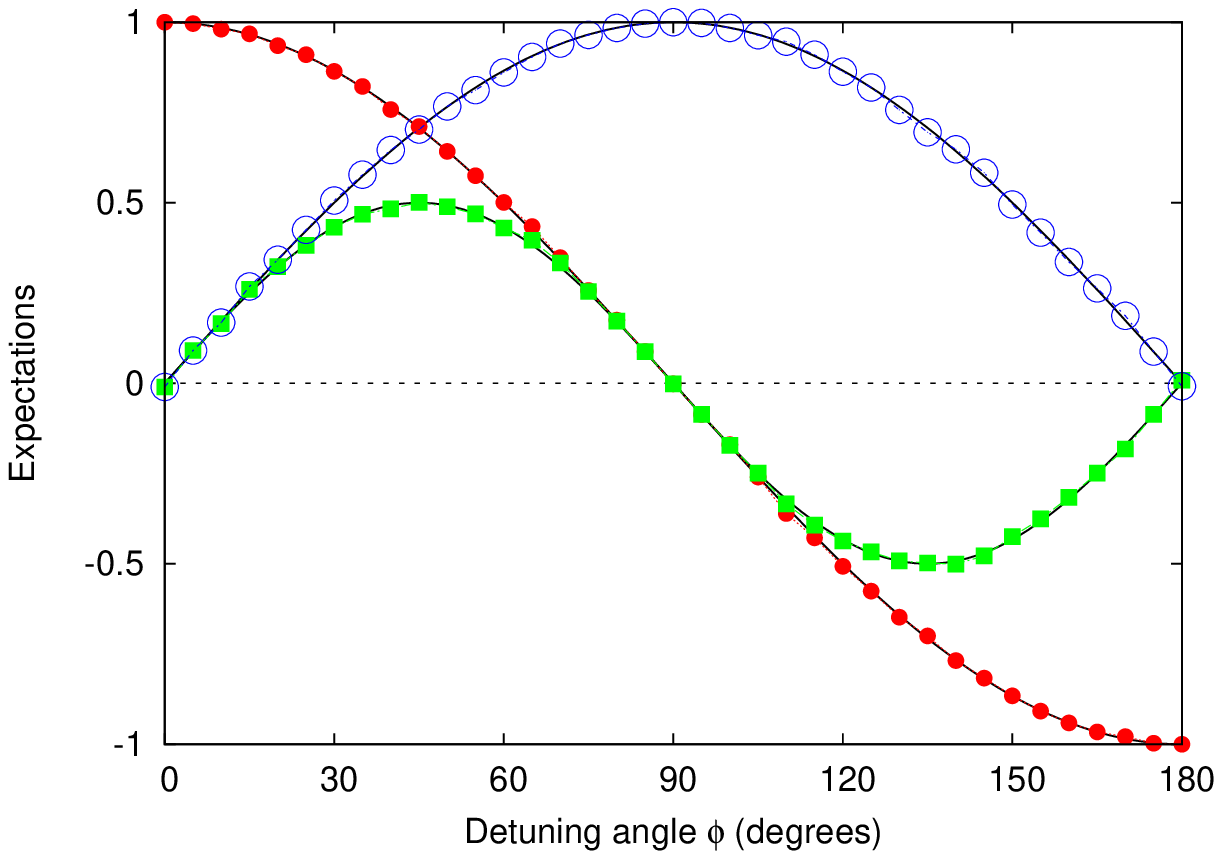}
\includegraphics[width=8cm]{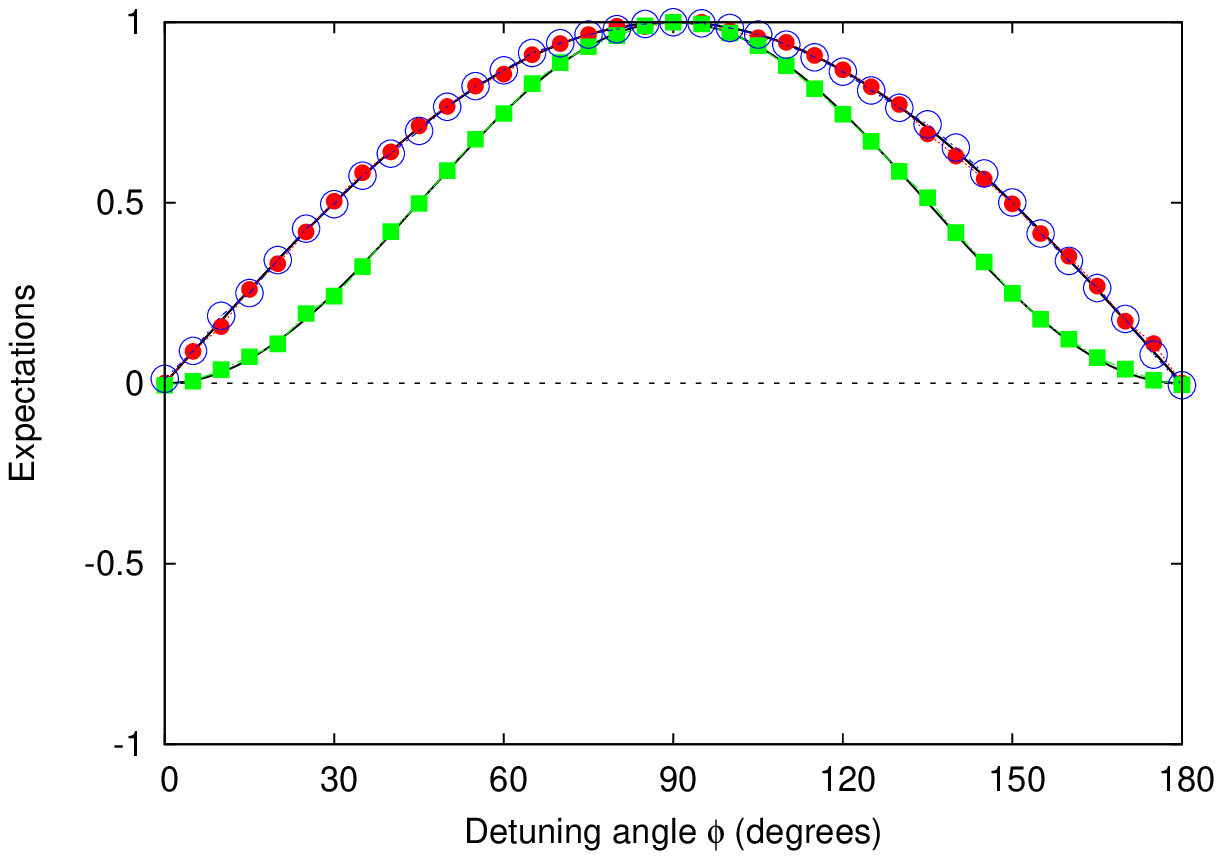}
\end{center}
\caption{Results for the expectations
$\langle S_1\rangle$ (red solid circles),
$\langle S_2\rangle$ (green solid squares), and
$\langle S_1S_2\rangle$ (blue open circles)
as obtained by the event-by-event simulation of the neutron experiment~\citep{ERHA12,SULY13},
using the model Eq.~(\ref{eve4}) for the spin analyzers.
The solid lines represent the corresponding quantum theoretical prediction
as obtained from Eq.~(\ref{qt8}).
Left: incoming particles have magnetization $(1,0,0)$.
Right: incoming particles have magnetization $(0,1,0)$.
For each pair of settings $(S_1,S_2)$ of the spin analyzers (SA2,SA3)
and each position of the pair of spin flippers (SF2,SF3) represented by a rotation of $\phi$ about the $z$-axis,
referred to as detuning angle in \citep{ERHA12,SULY13}, the simulation consists of sending
$N=10000$ messengers (``neutrons'') into stage 2.
}
\label{res1}
\end{figure}

{\color{black}

\begin{figure}[t]
\begin{center}
\includegraphics[width=8cm]{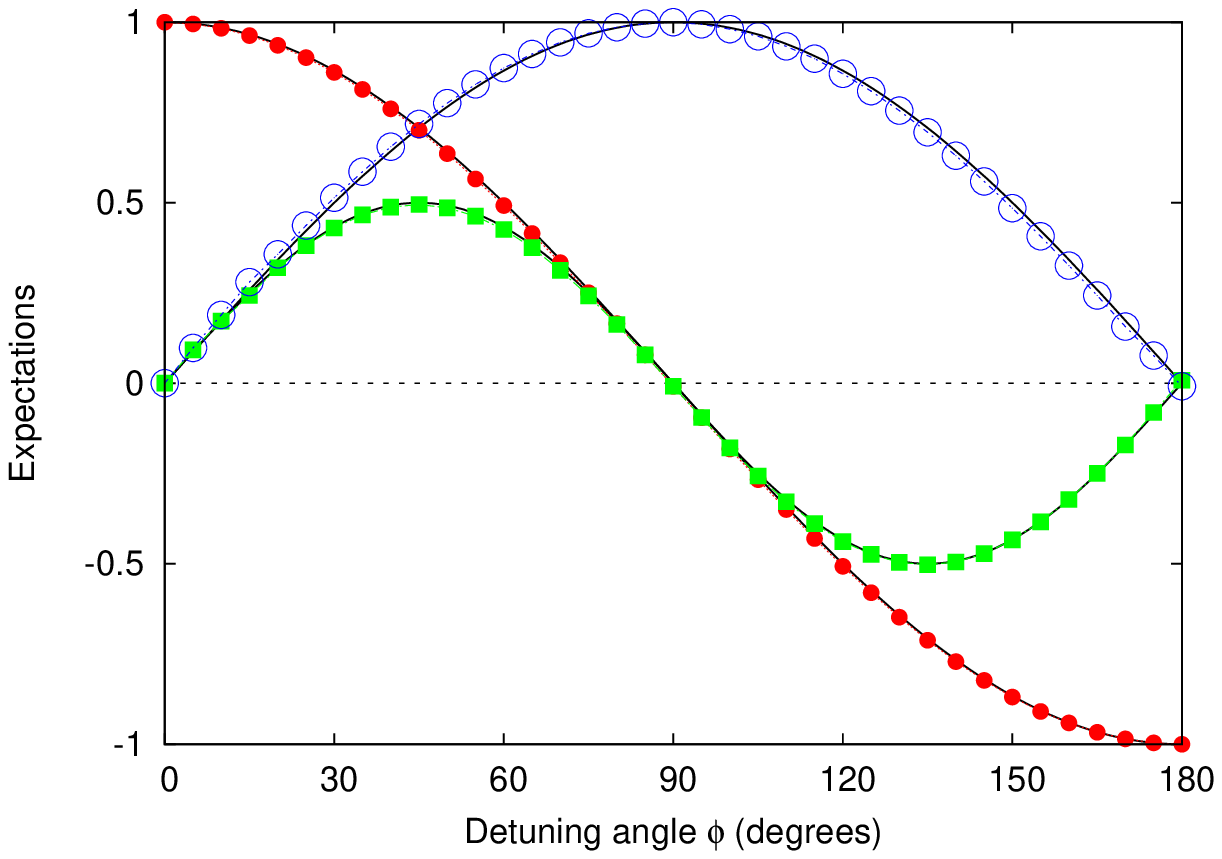}
\includegraphics[width=8cm]{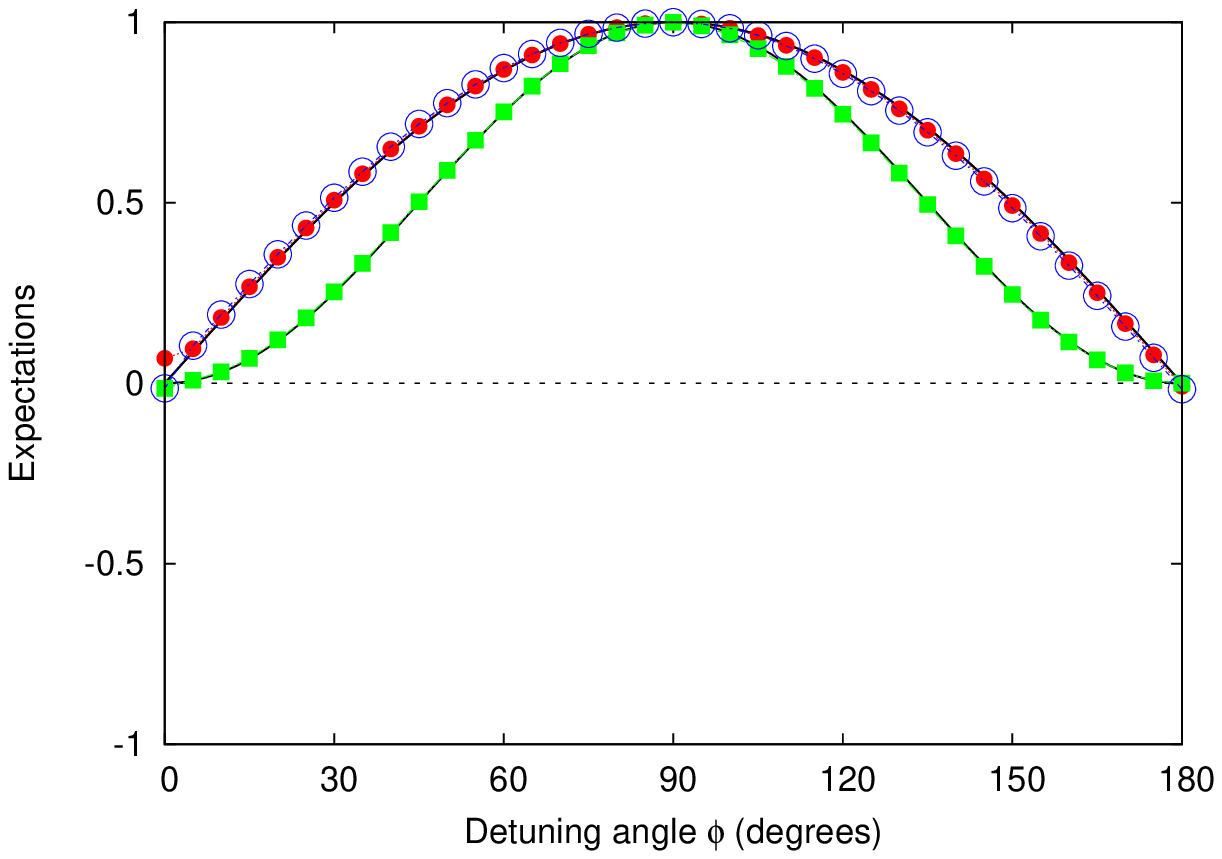}
\end{center}
\caption{Same as Fig.~\ref{res1}
except that the probabilistic model for the spin-analyzer, as specified by Eq.~(\ref{eve4}),
has been replaced by a deterministic learning machine with an update rule given by Eq.~(\ref{eve4a}).
For each pair of settings $(S_1,S_2)$ of the spin analyzers (SA2,SA3)
and each position of the pair of spin flippers (SF2,SF3) represented by a rotation of $\phi$ about the $z$-axis,
referred to as detuning angle in \citep{ERHA12,SULY13}, the simulation consists of sending
$N=10000$ messengers (``neutrons'') into stage 2.
The parameter $\gamma=0.999$.
}
\label{res1a}
\end{figure}
}

Summarizing: the event-based simulation model of the neutron experiment~\citep{ERHA12,SULY13}
presented in this section does not rely, in any sense, on concepts of quantum theory yet it reproduces
all features of the quantum theoretical description of the experiment.
Although the event-based model is classical in nature, it is not classical in the sense that
it cannot be described by classical Hamiltonian dynamics.

\section{Uncertainty relations: theory}

The neutron experiment~\citep{ERHA12,SULY13} was conceived to test an error-disturbance uncertainty relation
proposed by Ozawa~\citep{OZAW03}.
By introducing particular definitions of the measurement error $\epsilon(A)$
of an operator $A$ and the disturbance $\eta(B)$ of an operator $B$, Ozawa showed that
\begin{eqnarray}
\epsilon(A)\eta(B)+
\epsilon(A)\Delta(B)+
\Delta(A)\eta(B) &\ge& \frac{1}{2} |\langle [A,B]\rangle|
,
\label{unc0}
\end{eqnarray}
where
\begin{eqnarray}
\epsilon^2(A)&=&\langle (M_A-A)^2\rangle,
\label{unc1a}
\\
\eta^2(B)&=&\langle (M_B-B)^2\rangle,
\label{unc1b}
\\
\Delta^2(A)&=&\langle A^2\rangle - \langle A\rangle^2,
\label{unc1c}
\\
\Delta^2(B)&=&\langle B^2\rangle - \langle B\rangle^2
,
\label{unc1d}
\end{eqnarray}
and $M_A$ and $M_B$ represent the operators of different measuring devices
(implying $[M_A,M_B]=0$) that allow us to read
of the value of the measurement of $A$ and $B$, respectively.
Thereby it is assumed that
$\langle M_A\rangle = \langle A\rangle$ and
$\langle M_B\rangle = \langle B\rangle$, that is
that the measurements of $A$ and $B$ are unbiased, implying
that $\epsilon(A)=\Delta(M_A-A)$ and $\eta(B)=\Delta(M_B-B)$.

In Ozawa's model of the measurement process, the state of the system + measurement devices is represented
by a direct product of the wavefunction of the system
and the wavefunction of the measurement devices~\citep{OZAW03}.
The operators $A$ and $B$ refer to the dynamical variables of the system
while the $M_A$ and $M_B$ refer to the dynamical variables
of two different measurement devices.
Furthermore, it is assumed that both the system
and the measuring devices (probes) are described by quantum theory, i.e.
the time evolution of the whole system is unitary~\citep{OZAW03,FUJI12}.
Although this basic premise is at odds with the fact that experiments
yield definite answers~\citep{BALL70,LEGG87,HOME97,BALL03},
within the realm of the quantum theoretical model, it ``defines'' the measurement process,
see~\citep{NIEU13} for an extensive review.

Following~\citep{FUJI12,FUJI13a}, inequalities such as Eq.~(\ref{unc0}) are readily derived
by starting from the identity
\begin{eqnarray}
[C-A,D-B]&=&[C,D]-[A,D]-[C,B]+[A,B]
\nonumber \\&=&
[C,D]-\left([A,D-B]+[A,B]\right) - \left([C-A,B]+[A,B]\right)+[A,B]
\nonumber \\&=&
[C,D]-[A,D-B] -[C-A,B]-[A,B]
.
\label{unc4}
\end{eqnarray}
Assuming that $[C,D]=0$, we have
\begin{eqnarray}
[C-A,D-B]+[A,D-B]+[C-A,B]&=&-[A,B]
,
\label{unc5}
\end{eqnarray}
or, taking expectation values,
\begin{eqnarray}
\langle[C-A,D-B]\rangle+\langle[A,D-B]\rangle+\langle[C-A,B]\rangle&=&-\langle[A,B]\rangle
.
\label{unc6}
\end{eqnarray}
Taking the absolute value of both sides of Eq.~(\ref{unc6}) and using the triangle inequality we find
\begin{eqnarray}
|\langle[C-A,D-B]\rangle|+|\langle[A,D-B]\rangle|+|\langle[C-A,B]\rangle|&\ge&|\langle[A,B]\rangle|
.
\label{unc7}
\end{eqnarray}
Next, we apply the inequality $2\Delta(X)\Delta(Y)\ge|\langle [X,Y]\rangle|$ ~\citep{ROBE29,BALL03}
to each of the three terms in Eq.~(\ref{unc7}) and obtain
\begin{eqnarray}
\Delta(C-A)\Delta(D-B)+\Delta(A)\Delta(D-B)+\Delta(C-A)\Delta(B)&\ge&\frac{1}{2}|\langle[A,B]\rangle|
.
\label{unc8}
\end{eqnarray}
The derivation of Eq.~(\ref{unc8}) only makes use of the triangle inequality, the notion
of a non-negative inner product on a vector space, the Cauchy-Schwarz inequality
and the assumption that $[C,D]=0$.
Therefore Eq.~(\ref{unc8}) is ``universally valid''~\citep{OZAW03,FUJI12,FUJI13a} whenever $[C,D]=0$.

Inequality Eq.~(\ref{unc0}) directly follows from Eq.~(\ref{unc8}) by
substituting $C=M_A$, $D=M_B$ and by using the assumption of unbiasedness,
meaning $\langle M_A - A\rangle =0$ and $\langle M_B - B\rangle=0$.
With the restriction imposed by the assumptions of unbiasedness and $[M_A,M_B]=0$,
it is also ``universally valid''.

In contrast, the common interpretation of Heisenberg's original writings~\citep{HEIS27} suggests
an uncertainty relation which reads~\citep{OZAW03,ERHA12,SULY13,FUJI12,FUJI13a}
\begin{eqnarray}
\epsilon(A)\eta(B) &\ge& \frac{1}{2} |\langle [A,B]\rangle|
.
\label{unc2}
\end{eqnarray}
Thereby it is assumed, without solid justification,
that $\epsilon(A)$ and $\eta(B)$ correspond to the ``uncertainties'' which Heisenberg had in mind,
see also ~\citep{BUSH07,BUSH13}.

Unlike Eq.~(\ref{unc0}), inequality Eq.~(\ref{unc2})
lacks a mathematical rigorous basis and therefore it is not a surprise that it can be violated~\citep{BALL70}.
Indeed, the data recorded in the neutron experiment clearly violate Eq.~(\ref{unc2})~\citep{ERHA12,SULY13}.
In general, in  mathematical probability theory as well as quantum theory,
inequalities such as the Cram\'er-Rao bound~\citep{TREE68}, the Robertson inequality~\citep{ROBE29}, Eq.~(\ref{unc0}) and Eq.~(\ref{unc8})
are mathematical identities which result from applications of the Cauchy-Schwarz inequality.
Being mathematical identities within the realm of standard arithmetic, they are void of any physical meaning
and cannot be violated.
Therefore, if an experiment indicates that such an identity (i.e. inequality) might be violated,
this can only imply that there is an ambiguity (error) in the mapping between the variables
used in the theoretical model and those assigned to the experimental observations~\citep{BO1862,RAED11a}.
Any other conclusion that is drawn from such a violation cannot be justified on logical/mathematical grounds.

Following~\citep{ERHA12,SULY13}, we assume that the state of the system is represented by
the density matrix $\rho=|\mathbf{z}\rangle \langle\mathbf{z}|$, that is
the magnetic moment of the neutrons are assumed to be aligned along the $z$-direction.
With $A=\sigma_x$ and $B=\sigma_y$ we have~\citep{ERHA12,SULY13}
\begin{eqnarray}
M_A&=&\sigma_\phi=\sigma_x\cos\phi+\sigma_y\sin\phi,
\nonumber \\
M_B&=&\frac{1}{4}\left( (1+\sigma_\phi)\sigma_y(1+\sigma_\phi)+(1-\sigma_\phi)\sigma_y(1-\sigma_\phi)\right)=\sigma_\phi \sin\phi=\sin\phi M_A,
\nonumber \\
\epsilon^2(A)&=&\langle (\sigma_\phi-\sigma_x)^2\rangle
=2 \langle \openone -\cos\phi \sigma_x^2\rangle
=
4\sin^2\frac{\phi}{2}
\nonumber \\
\eta^2(B)&=&
\langle [\sigma_\phi,\sigma_y]^\dagger [\sigma_\phi,\sigma_y]\rangle/2
=\cos^2\phi \langle [\sigma_x,\sigma_y]^\dagger [\sigma_x,\sigma_y]\rangle/2
=2\cos^2\phi\langle \sigma_z^2\rangle
=
2\cos^2\phi
,
\label{unc3}
\end{eqnarray}
and $\sigma(A)=\sigma(B)=1$.
Combining Eq.~(\ref{unc0}) and Eq.~(\ref{unc3}) yields~\citep{ERHA12,SULY13}
\begin{eqnarray}
\epsilon(A)\eta(B)+
\epsilon(A)\sigma(B)+
\sigma(A)\eta(B)
= 2\sqrt{2}\cos\phi\sin\frac{\phi}{2}+2\sin\frac{\phi}{2}+\sqrt{2}\cos\phi\ge1
.
\label{unc9}
\end{eqnarray}
{\color{black}
Note the absence of $\hbar$ in Eq.~(\ref{unc9}),
in agreement with work that shows that $\hbar$
may be eliminated from the basic equations of (low-energy) physics
by a re-definition of the units of mass, time, etc.~\citep{VOLO10,RALS13a}.
}

Conceptually, the application of Eq.~(\ref{unc0}) to the neutron experiment~\citep{ERHA12,SULY13} is not
as straightforward as it may seem.
In a strict sense, in the neutron experiment~\citep{ERHA12,SULY13}, there are no measurements of the kind envisaged
in Ozawa's measurement model.
This is most obvious from the quantum theoretical description of the experiment given in Section~\ref{sec2}:
for fixed $S_1$ and $S_2$, the relative frequency of detector counts is given by Eq.~(\ref{qt8}),
and ``noise'' caused by ``probes'' does not enter the description.
Indeed, from the expressions of $\epsilon^2(A)$ and $\eta^2(B)$ in terms of spin operators, see Eq.~(\ref{unc3}),
it is immediately clear that in order to determine $\epsilon^2(A)$ and $\eta^2(B)$,
there is no need to actually measure a dynamical variable.
Moreover, in the laboratory experiment,
the values of $S_1$ and $S_2$ are not actually measured
but, as they represent the orientation of the spin analyzers SA2 and SA3,
are kept fixed for a certain period of time.
Unlike in the thought experiment for which Eq.~(\ref{unc0}) was derived,
the outcome of an experimental run is not the set of pairs $(S_1,S_2)$ but rather
the number of counts for this particular set of settings.
Nevertheless, with some clever manipulations~\citep{ERHA12,SULY13}, it is possible to express
the unit operators that appear in Eq.~(\ref{unc3}) in terms of dynamical variables, the
expectations of which can be extracted from the data of single-neutron experiments.

If the state of the spin-1/2 system
is described by the density matrix $\rho=|\mathbf{z}\rangle \langle\mathbf{z}|$, we have~\citep{ERHA12,SULY13}
\begin{eqnarray}
\epsilon^2(A)&=&2+\langle\mathbf{z}|M_A  |\mathbf{z}\rangle
+\langle\mathbf{-z}| M_A |\mathbf{-z}\rangle
-2\langle\mathbf{x}| M_A |\mathbf{x}\rangle
\nonumber \\&=&
2-2\sum_{S_1,S_2=\pm1} S_1 P(S_1,S_2|\mathbf{a}=(1,0,0))
\nonumber \\
\eta^2(B)&=&2+\langle\mathbf{z}| M_B |\mathbf{z}\rangle
+\langle\mathbf{-z}| M_B |\mathbf{-z}\rangle
-2\langle\mathbf{y}| M_B |\mathbf{y}\rangle
\nonumber \\&=&
2-2\sum_{S_1,S_2=\pm1} S_2 P(S_1,S_2|\mathbf{a}=(0,1,0))
,
\label{unc10}
\end{eqnarray}
where we used $\langle\mathbf{\pm z}| M_A |\mathbf{\pm z}\rangle=\langle\mathbf{\pm z}| M_B |\mathbf{\pm z}\rangle=0$
and $P(S_1,S_2|\mathbf{a})$ is given by Eq.~(\ref{qt8}).

The expressions Eq.~(\ref{unc10}) are remarkable: they show that
$\epsilon^2(A)$ and $\eta^2(B)$ have to be obtained from two incompatible experiments,
namely with initial magnetic moments along $\mathbf{x}$ and $\mathbf{y}$, respectively.
From the point of view of probability theory, this immediately raises the question
why, in this particular case,
it is possible to derive mathematically meaningful results that involve
two different conditional probability distributions with incompatible conditions.
As first pointed out by Boole~\citep{BO1862} and generalized by Vorob'ev~\citep{VORO62}, this
is possible if and only if there exists a ``master'' probability distribution for the
union of all the incompatible conditions.
For instance, in two- and three-slit experiments~\citep{BALL86,BALL03,SINH10,RAED12c}
such a master probability distribution does not exist by construction of the experiment.
Another prominent example is the violation of
one or more Bell inequalities which is known to be mathematically equivalent
to the statement that a master probability distribution for the relevant combination of experiments
does not exist~\citep{BO1862,FINE82a,HESS01b,RAED11a}.
However, in contrast to these two examples,
in the case of the neutron experiment, one can devise a realizable experiment that
simultaneously yields all the averages that can be obtained from
two experiments (one with $\mathbf{a}=\mathbf{x}$ and another one with $\mathbf{a}=\mathbf{y}$)
of the kind shown in Fig.~\ref{stages} or Fig.~\ref{fig0}.

Our proof is based on the extension of the filtering-type experiment shown in Fig.~\ref{fig0}
to three dichotomic variables.
Imagine that instead of placing detectors in the output beams that emerge from magnets
$M_1$ and $M_2$, we place four identical magnets with their magnetic fields
in the direction $\mathbf{d}$ and count the particles in each of the eight beams.
A calculation, similar to the one that lead to Eq.~(\ref{fil6}), yields~\citep{RAED11a}
\begin{eqnarray}
P(S_{1},S_{2},S_{3}|\mathbf{a})&=&
\mathbf{Tr} \rho M(S_1,\mathbf{b})M(S_2,{\mathbf c})M(S_3,{\mathbf d})M(S_2,{\mathbf c})M(S_1,{\mathbf b})
\nonumber \\&=&
\frac{1+ \mathbf{a}\cdot\mathbf{b}\;S_1+ \mathbf{a}\cdot\mathbf{b}\;\mathbf{b}\cdot\mathbf{c}\;S_2
+ \mathbf{a}\cdot\mathbf{b}\;\mathbf{b}\cdot\mathbf{c}\;\mathbf{c}\cdot\mathbf{d}\;S_3
+\mathbf{a}\cdot\mathbf{b}\;\mathbf{c}\cdot\mathbf{d}\;S_1S_2 S_3
}{8}
\nonumber \\
&&+\frac{
\mathbf{b}\cdot\mathbf{c}\;S_1 S_2
+\mathbf{b}\cdot\mathbf{c}\;\mathbf{c}\cdot\mathbf{d}\;S_1 S_3
+\mathbf{c}\cdot\mathbf{d}\; S_2 S_3
}{8}
,
\label{fil6a}
\end{eqnarray}
for the probability to observe the given triple $(S_{1},S_{2},S_{3})$.
Choosing $\mathbf{a}=\mathbf{x}$, $\mathbf{c}=\mathbf{y}$, and $\mathbf{b}=\mathbf{d}=\mathbf{x}\cos\phi+\mathbf{y}\sin\phi$
it follows immediately that $\langle S_1\rangle$, $\langle S_2\rangle$, and
$\langle S_1S_2\rangle$ obtained from Eq.~(\ref{fil6}) with $\mathbf{a}=\mathbf{x}$
agree with the same averages computed from Eq.~(\ref{fil6a}).
Likewise, $\langle S_1\rangle$, $\langle S_2\rangle$, and
$\langle S_1S_2\rangle$ obtained from Eq.~(\ref{fil6}) with $\mathbf{a}=\mathbf{y}$
coincide with $\langle S_2S_3\rangle$, $\langle S_1S_3\rangle$ and $\langle S_1S_2\rangle$
computed from Eq.~(\ref{fil6a}).

\section{Uncertainty relations: event-based simulation}

In the neutron experiment~\citep{ERHA12,SULY13} and therefore also in our event-based simulation,
the numerical values of $\epsilon(A)$ and $\eta(B)$ are obtained by counting detection events.
Let $N(S_1,S_2|\mathbf{a})$ denote the count for the case
in which the direction of the magnetic moment of the incoming neutrons (after stage 1) is
$\mathbf{a}$ and
the analyzers SA2 and SA3 are along the directions $S_1$ and $S_2$, respectively.
Then, we have
\begin{eqnarray}
\epsilon^2(A)&\approx&
2-2\frac{\sum_{S_1,S_2=\pm1} S_1 N(S_1,S_2|\mathbf{x})}{\sum_{S_1,S_2=\pm1} N(S_1,S_2|\mathbf{x})}
\nonumber \\
\eta^2(B)&\approx&
2-2\frac{\sum_{S_1,S_2=\pm1} S_2 N(S_1,S_2|\mathbf{y})}{\sum_{S_1,S_2=\pm1} N(S_1,S_2|\mathbf{y})}
.
\label{unc10a}
\end{eqnarray}

\begin{figure}[t]
\begin{center}
\includegraphics[width=8cm]{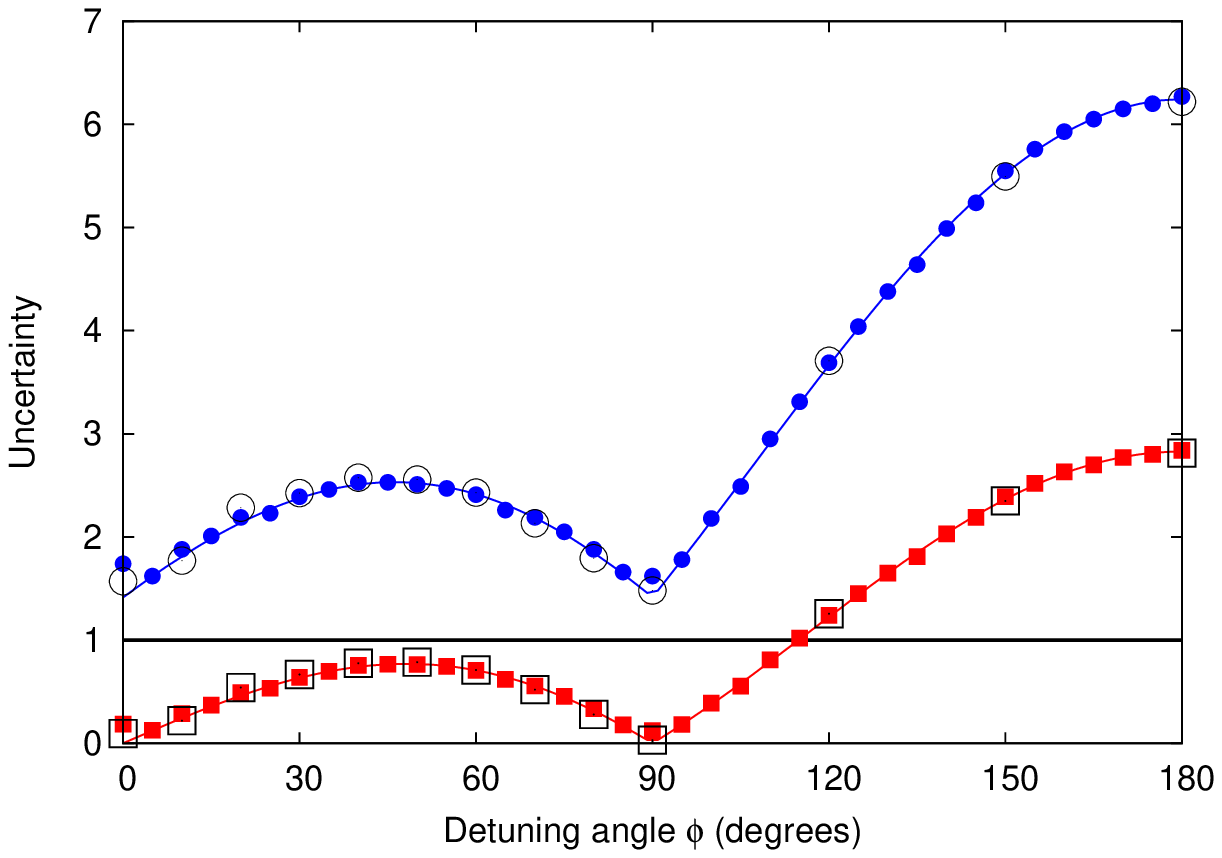}
\includegraphics[width=8cm]{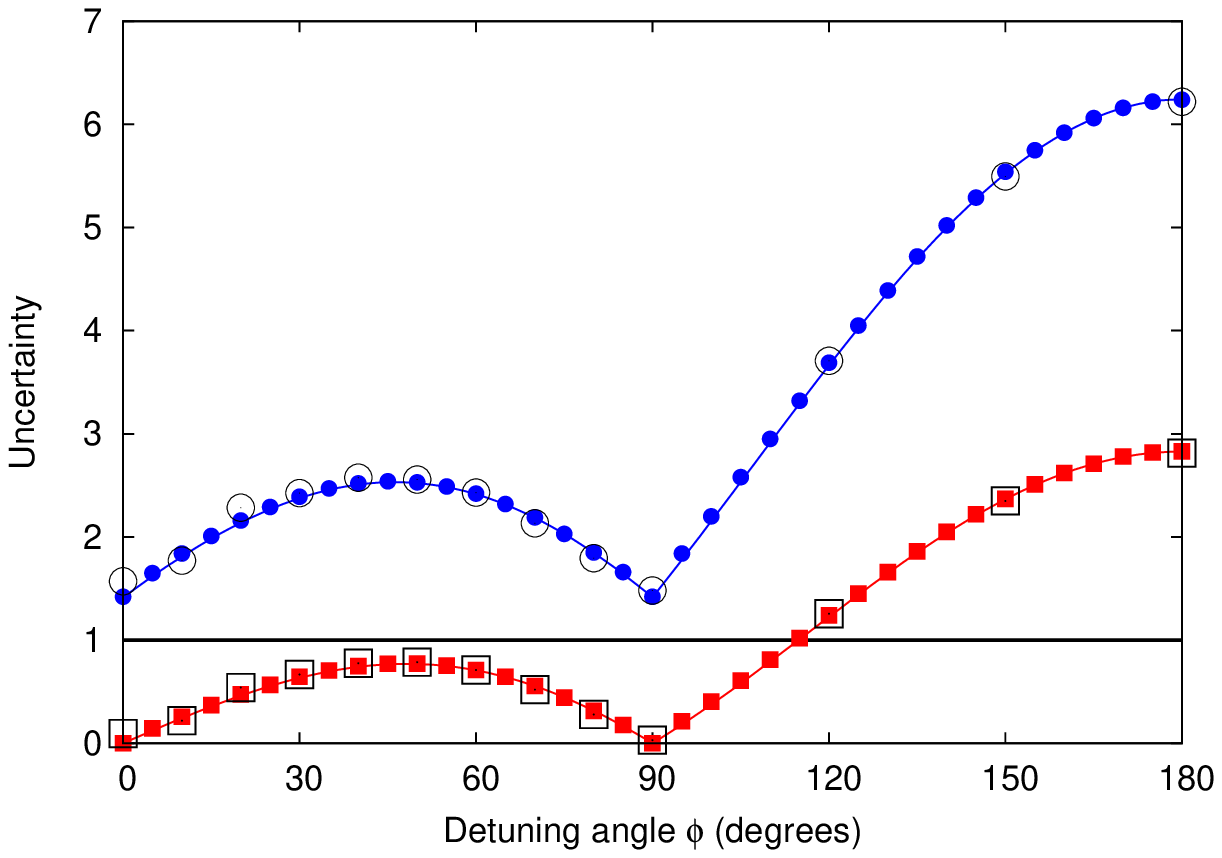}
\end{center}
\caption{%
{\color{black}
Uncertainties
$\epsilon(A)\eta(B)+\epsilon(A)\sigma(B)+\sigma(A)\eta(B)$ (circles)
and
$\epsilon(A)\eta(B)$ (squares)
as obtained from the event-by-event simulations (solid symbols) of the single-neutron experiment~\citep{ERHA12,SULY13},
the experiment itself ((open symbols), data kindly provided to us by G. Sulyok and Y. Hasagawa)
and quantum theory.
The solid horizontal line represents the lowerbound in Eq.~(\ref{unc0}).
The other solid lines represent the corresponding quantum theoretical prediction
as obtained from Eq.~(\ref{unc3}).
It is clear that the naive application of the Heisenberg uncertainty relation,
$\epsilon(A)\eta(B)\ge1$ is at odds with the prediction of quantum theory,
the event-based simulation data and with the experimental data~\citep{ERHA12,SULY13}.
On the other hand, the experimental data (open symbols, ~\citep{ERHA12,SULY13})
and the results of the event-based simulation (solid symbols)
comply with inequality Eq.~(\ref{unc0}).
Left: Simulation performed using the probabilistic model for the spin analyzer, see Eq.~(\ref{eve4}).
Right: Simulation performed using the deterministic learning machine model for the spin analyzer, see Eq.~(\ref{eve4a}),
with $\gamma=0.999$.
In both cases, the number of input events per detuning angle is $N=10000$.
}
}
\label{res2}
\end{figure}

As shown in~\citep{ERHA12,SULY13}, the neutron counts observed in the single-neutron experiment
yield numerical values of $\epsilon(A)\eta(B)+\epsilon(A)\sigma(B)+\sigma(A)\eta(B)$
which are in excellent agreement with the quantum theoretical prediction
$2\sqrt{2}\cos\phi\sin(\phi/2)+2\sin(\phi/2)+\sqrt{2}\cos\phi$.

We have already demonstrated that the ``classical'' event-based simulation model produces
results for the averages which, within the statistical errors,
cannot be distinguished from those predicted by quantum theory.
Therefore, it is to be expected that the data generated by the event-by-event simulation
also satisfies the universally valid error-disturbance uncertainty relation Eq.~(\ref{unc0})
and as shown in Fig.~\ref{res2}, this is indeed the case.
As expected, the data produced by the event-based simulation also violate Eq.~(\ref{unc2}),
{\color{black}
independent of whether we use pseudo-random numbers (see Eq.~(\ref{eve4}))
or the DLM rule (see Eq.~(\ref{eve4a})) to model the operation of the spin analyzer.
}

\begin{figure}[t]
\begin{center}
\includegraphics[width=12cm]{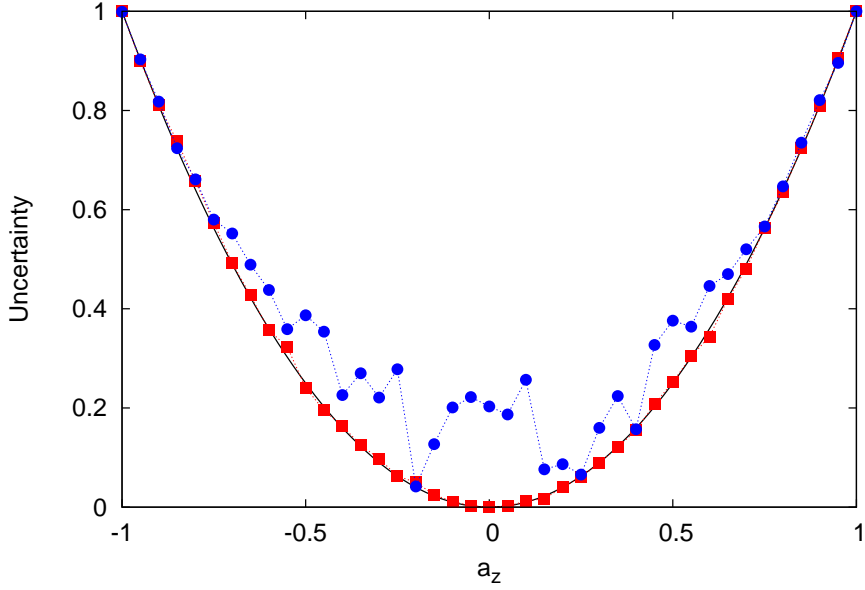}
\end{center}
\caption{Event-based simulation results for
$(1-\langle \sigma_x\rangle^2)(1-\langle \sigma_y\rangle^2)$ (blue circles)
and
$\langle \sigma^z\rangle^2$ (red squares)
for different values of $-1\le a_z\le 1$.
The solid black line represents the quantum theoretical lowerbound $a_z^2$.
For each value of $a_z$, the initial direction
of the magnetic moments is $(a_x,a_y,a_z)$ where
$(a_x,a_y)$ is a point on the circle with radius $\sqrt{1-a_z^2}$
chosen using a uniform pseudo-random number.
The fluctuations of the data
$(1-\langle \sigma_x\rangle^2)(1-\langle \sigma_y\rangle^2)$
reflect the fact that the initial states with different values of $a_z$
are uncorrelated.
For each value of $a_z$, $N=10000$ messengers were created.
The dotted line is a guide to the eyes only.
}
\label{res3}
\end{figure}

Finally, for the sake of completeness, we show that the event-by-event simulation produces data which complies with
the standard Heisenberg-Robertson uncertainty relation $\Delta(\sigma_x)\Delta(\sigma_y)\ge |\langle \sigma_z \rangle|$.
Without loss of generality, the state of the spin-1/2 particle
may be represented by the density matrix Eq.~(\ref{qt6}), also if it is interacting with other degrees of freedom
and the inequality $\Delta(\sigma_x)\Delta(\sigma_y)\ge |\langle \sigma_z \rangle|$ reads
\begin{eqnarray}
(1-\langle \sigma_x\rangle^2)(1-\langle \sigma_y\rangle^2)&\ge& \langle \sigma_z\rangle^2
,
\label{com0}
\\
\noalign{or, using Eq.~(\ref{qt6}),}
\nonumber \\
(1-a_x^2)(1-a_y^2)&\ge& a_z^2
.
\label{com1}
\end{eqnarray}
The last inequality also trivially follows from the constraint $a_x^2+a_y^2+a_z^2\le1$.
{\color{black}
As in the case of Eq.~(\ref{unc9}), there is no $\hbar$ in Eq.~(\ref{com0}),
in agreement with the idea that $\hbar$
may be eliminated by re-defining the units of mass, time, etc.~\citep{VOLO10,RALS13a}.
}

The simulation procedure that we use is as follows.
\begin{enumerate}[\ \bf 1.\ ]
\item
Loop over the values $(\mathbf{m}_{\mathrm{initial}})_z=a_z=-1,\ldots 1$ in small steps, e.g. in steps
of 0.05.
\item
Generate a uniform pseudo-random number $0<r<1$.
Compute $(\mathbf{m}_{\mathrm{initial}})_x=a_x=\sqrt{1-a_z^2}\cos(2\pi r)$
and $(\mathbf{m}_{\mathrm{initial}})_y=a_y=\sqrt{1-a_z^2}\sin(2\pi r)$.
This step yields a direction of the magnetization $\mathbf{m}_{\mathrm{initial}}$
which is chosen randomly in the $x-y$ plane.
\item
Generate $N$ messengers with message $\mathbf{m}_{\mathrm{initial}}$
and send them through a spin analyzer aligned along the $x$-direction.
Count the messengers that pass the spin analyzer.
Repeat this procedure for spin analyzers aligned along the $-x$,
$\pm y$ and $\pm z$-direction.
Processing the $N$ messengers yields the
counts $N(\mathbf{x}|\mathbf{a})$, $N(\mathbf{-x}|\mathbf{a})$, etc.
\item
Compute the averages
$\langle \sigma_x\rangle\approx(N(\mathbf{x}|\mathbf{a})-N(\mathbf{-x}|\mathbf{a}))/(N(\mathbf{x}|\mathbf{a})+N(\mathbf{-x}|\mathbf{a}))$,
etc.
\item
Go to step {\bf 1.} as long as $a_z\le1$.
\item
Plot the results for $(1-\langle \sigma_x\rangle^2)(1-\langle \sigma_y\rangle^2)$
and $\langle \sigma^z\rangle^2$ as a function of $a_z$.
\end{enumerate}

The results of the event-based simulation are shown in Fig.~\ref{res3}.
Within the usual statistical errors, the classical, statistical model
produces data which comply with the Heisenberg-Robertson uncertainty relation Eq.~(\ref{com0}).

\section{Discussion}

We have shown that a genuine classical event-based model can produce events such that their statistics
satisfies the (generalized) Heisenberg-Robertson uncertainty relation which,
according to present teaching, is a manifestation of truly quantum mechanical behavior.

One might be tempted to argue that in the event-based model, the direction of magnetic moment
is known exactly and can therefore not be subject to uncertainty.
However, this argument is incorrect in that it ignores the fact that the model of the spin analyzers
{\color{black}
generates
(through the use of pseudo-random numbers, see Eq.~(\ref{eve4}) or the update rule Eq.~(\ref{eve4a}))
a distribution of outcome frequencies.
}
In fact, as is well-known, the variance of any statistical experiment 
(including those that are interpreted in terms of quantum theory) satisfies the
Cram\'er-Rao bound, a lower bound on the variance of estimators of a parameter
of the probability distribution in terms of the Fisher information~\citep{TREE68}.
The Cram\'er-Rao bound contains, as a special case, Robertson's inequality
$\Delta(x)\Delta(p)\ge \hbar/2$~\citep{STAM59,FRIE89,FRIE04,KAPS10,KAPS11,SKAL11}.
The observation that a classical statistical model produces data that complies with
``quantum theoretical'' uncertainty relations is a manifestation of this general mathematical result.
The uncertainty relations provide bounds on the statistical uncertainties in the data and,
as shown by our event-based simulation of the neutron experiment~\citep{ERHA12,SULY13},
are not necessarily a signature of quantum physics, conjugate variables, etc.

{\color{black}
As mentioned in the introduction, the event-based approach has successfully been applied
to a large variety of single-photon and single-neutron experiments that involve
interference and entanglement. In the present paper, we have shown
that, without any modification, the same simulation approach can also mimic,
event-by-event, an experiment that probes ``quantum  uncertainty''.
As none of these demonstrations rely on concepts of quantum theory
and as it is unlikely that the success of all these demonstrations is accidental,
one may wonder what it is that makes a system genuine ``quantum''.

In essence, in our work we adopt Bohr's point of view
that ``There is no quantum world. There is only an abstract physical description''
(reported by~\citep{PETE63}, for a discussion see~\citep{PLOT10a})
and that ``The physical content of quantum mechanics is exhausted
by its power to formulate statistical laws''~\citep{BOHR99}.
Or, to say it differently,
quantum theory describes our knowledge of the atomic phenomena rather than
the atomic phenomena themselves~\citep{LAUR97}.
In other words, our viewpoint is that quantum theory captures,
and does so extremely well, the inferences that we, humans,
make on the basis of experimental data~\citep{RAED13z}.
However it does not describe cause-and-effect processes.
Quantum theory predicts the probabilities that events occur,
but it cannot answer the question ``Why are there events?''~\citep{ENGL13},
very much as Euclidean geometry cannot answer the question ``What is a point?''.
On a basic level, it is our perceptual and cognitive system that defines, registers and processes events.
Events and the rules that create new events are the key elements of the event-based approach.
There is no underlying theory that is supposed to give rise to events
and everything follows by inference on the basis of the generated data,
very much like in real experiments.

The implication of the work presented in our paper
is that the beautiful single-neutron experiments~\citep{ERHA12,SULY13}
can be explained in terms of cause-and-effect processes in an event-by-event manner,
without reference to quantum theory and on a level of detail about which quantum theory
has nothing to say.
Furthermore, our work suggests that the relevance of ``quantum theoretical'' uncertainty relations
to real experiments needs to be reconsidered.
}

\section*{Disclosure/Conflict-of-Interest Statement}
The authors declare that the research was conducted in the absence of any commercial or financial relationships
that could be construed as a potential conflict of interest.

\section*{Acknowledgement}
We thank Koen De Raedt, Karl Hess, Thomas Lippert and Seiji Miyashita for many stimulating discussions.

\bibliography{c:/d/papers/all13}

\end{document}